\documentclass[amstex]{article}
\usepackage{amsmath}

\def\bx{{\mathbf x}}
\def\cK{{\mathcal K}}

\def\Dir{\,\,{\raise.15ex\hbox{/}\mkern-12mu D}}
\def\a{\alpha}

\def\d{\delta}
\def\D{\Delta}
\def\g{\gamma}
\def\G{\Gamma}

\def\m{\mu}
\def\n{\nu}

\def\pa{\partial}

\def\({\left(}
\def\){\right)}
\def\[{\left[}
\def\]{\right]}

\begin{document}
\ \\

\begin{center}
{\Large{\bf Regular and irregular boundary conditions in AdS/CFT correspondence
for spinor field}}\\

\ \\

{\bf R.C.Rashkov}\footnote{
e-mail: rash@phys.uni-sofia.bg}  \\
\ \\
Department of Theoretical Physics\\
Sofia University, 5 J.Bourchier Blvd.\\
1164 Sofia, Bulgaria

\ \\

\end{center}

\ \\

\begin{abstract}
In a recent paper Klebanov and Witten proposed \cite{KW} to fomulate the AdS/CFT
correspondence principle by taking an "irregular boundary condition" for a
scalar field. In this paper we generalize this idea to the case of
spinor field with interaction. The action functional following from
the choice of irregular boundary conditions and which must be used in
the AdS/CFT correspondence is related to the usual action by a Legendre
transform. For the new theory we found the modified Green's function
that must be used for internal lines in calculating higher order graphs.
It is proved that the considerations are valid to all orders in
perturbation theory.
\end{abstract}

\section{Introduction}

The resent fascinating conjecture by Maldacena \cite{Malda} about the
correspondence between supergravity on d+1 Anti-de Sitter (AdS) space
and Conformal Field Theory (CFT) living on its asymptotic d-dimensional
boundary becomes a powerful tool in studying the String Theory.
The underlying principle behind this AdS/CFT
correspondence was elaborated in explicit form by Gubser, Klebanov and
Polyakov \cite{gkp} and Witten \cite{w}. According to \cite{gkp} and
\cite{w}, the action for the supergravity theory on AdS considered as a
functional of the asymptotic values of the fields on the boundary is
interpreted as a generating functional for the correlation functions in the
conformal field theory living on the boundary. The explicit form of this
interpretation is:
\begin{equation}
\int\limits_{\Phi}{\mathcal D}\Phi exp\{-S\[\Phi\]\}=
\langle exp\int\limits_{\pa AdS}d^d{\mathbf x}{\mathcal O}\Phi_0\rangle
\end{equation}
where $\Phi_0$ is the boundary data for AdS theory which couples to a
certain conformal operator $\mathcal O$ on the boundary. This
interpretation has already a number of proofs  for various cases of
fields form the fields content of supergravity theory on AdS space
\cite{w,5,6,HS,8,9,14,15,C,A,kr,Rash}\footnote{For a recent review see
\cite{OM}} . In all of the above particular
examples the field solutions can be splitted into two parts (see for
example \cite{Bala}) with different behavior near the boundary:
\begin{equation}
\Phi (x)\overset{x_0\to 0}{\approx} x_0^{\Delta_+}\Phi^++
\overset{x_0\to 0}{\approx} x_0^{\Delta_-}\Phi^-
\label{i0}
\end{equation}

where $\Phi^+$  are the fluctuating modes in the bulk and
$\Phi^-$  serves as a source for the correlation functions in the CFT living
on the boundary ($\Delta_+>\Delta_-$).Since the field
equations possess two solutions with different behavior, one must choose
one of them and
if the boundary conditions are imposed on the
fields $\Phi^-$ that are coupled on the
boundary to a conformal operator with larger dimension $\Delta$ such a
boundary conditions we will reffer as regular boundary conditions. Recently
Klebanov and Witten \cite{KW} observed that using this scheme of
calculations the unitary bound ($\frac d2-1\leq\Delta$) in the case of
scalar field cannot be achieved and they proposed an effective method to
include operators to do just that. The proposed method, generalized
later in \cite{Wi} to the interacting case, using the choice of
boundary conditions for the field solutions which would be coupled to
operators on the boundary with smaller dimension. In \cite{KW,Wi} was
shown that the respective boundary fields are cannonicaly conjugated to
each other and theyr actions are connected by a Legendre transform.
The fact that the corresponding boundary fields are conjugated was also
suggested in \cite{Vlado} by using the technique of intertwinning operators.
The action depending on the irregular boundary values serves as a
generating functional of the correlation functions. In \cite{KW,Wi} it was
found also the modified Green's function for this new theory that must
be used in the calculations of the correlation functions. The explicit
form of the two point correlation function in the case of scalar field
was found in \cite{KW,Wi}.

The aim of this article is to generalize the method of taking
irregular boundary conditions proposed for a scalar field in \cite{KW,Wi}
to the case of spinor field. As a byproduct of our considerations is
derivation of the modified Green's function for the spinor field that
must be used to calculate second and higher order graphs.

The structure of the article is as follows. In the rest of this Section
we will review the boundary term for a spinor fields along the lines of
\cite{H} and will comment briefly the possible boundary conditions. The
second Section is devoted to the results from AdS/CFT correspondence
principle connected to regular boundary conditions. In Section 3 we wll
consider a theory corresponding to irregular boundary conditions and the
two point correlation function will be computed. The correctness of
\cite{KW,Wi} proposal to all orders in perturbation theory will be
discussed.

Our starting point is the action for interacting spinor field given by:
\begin{equation}
S_D=\int\limits_{AdS}d^{d+1}x\sqrt{g}\bar\psi(x)\(\Dir -m\)\psi(x)+S_{int}
\label{i1}
\end{equation}
where $\sqrt{g}$ i determinat of the AdS metric.
We choose to work in coordinates
$x^a=(x^0,x^i)=(x^0,\vec x); i=1,\dots d$ defining $d+1$-dimensional
Euclidean Anti-de Sitter space as Lobachevski upper half plane $x^0>0$ with a
metric of the form:
\begin{equation}
ds^2=\frac{1}{x_0^2}\(dx^0+d{\vec x}^2\)
\label{2}
\end{equation}
With this choice the vielbein and the corresponding non-zero components
of the spin connection are given by the expressions:
\begin{equation}
e^\m_a=\frac{\d^a_\m}{x^0};\quad \omega_i^{0j}=-\omega_i^{j0}=
\frac{\d^j_i}{x^0};\quad a=0,\dots d
\label{3}
\end{equation}
The boundary of the AdS space consists in a hypersurface $x^0=0$ and a
single point $x^0=\infty$. The gamma matrices $\G^\m$ are connected to
the flat ones by $\G^\m=e^\m_a\g^a$ where $\g^a$ satisfy the usual
commutation relations $\{\g^a,\g^b\}=2\d^{ab}$.

In this frame the covariant derivative and the Dirac operator reads off:
\begin{equation}
D_\n=\pa_\n+\frac{1}{2x^0}\g_{0\n};\quad \G^\m D_\m=\Dir=x_0\g^0\pa_0+
x_0{\vec\g}.\vec\nabla-\frac d2\g^0
\label{4}
\end{equation}
where $\vec\g=(\g^i);\,\, \vec\nabla=(\pa_i);\,\, i=1\dots d$.

The equations of motion following from the action (\ref{i1}) are given
by:
\begin{equation}
\(\Dir-m\)\psi(x)=-\frac{\d S_{int}}{\d\bar\psi(x)}=-F(x)
\label{eq}
\end{equation}
\begin{equation}
\bar\psi(x)\(-\Dir-m\)=-\frac{\d S_{int}}{\d\psi(x)}=-\bar F(x)
\label{bareq}
\end{equation}
and the equation for the bulk-bulk Green's function for the above
equation is:
\begin{equation}
\(\Dir-m\)S(x;y)=S(x;y)\(-\Dir-m\)=\frac{1}{\sqrt{g}}\d^{d+1}(x-y)
\label{i3}
\end{equation}
with regulariry and boundary conditions:
$$
\lim_{x_0\to\infty}S(x;y)=\lim_{y_0\to 0}S(x;y)=0
$$
$$
\lim_{x_0\to 0}S(x;y)=\lim_{y_0\to\infty}S(x;y)=0
$$
The explicit solution of (\ref{i3}) will be presented later.
The solutions of equations (\ref{eq},\ref{bareq}) split into free field
solutions and interacting parts:
\begin{align}
\psi(x)&=\psi^{(h)}(x)-\int\limits_{AdS}d^{d+1}yS(x;y)F(y)
\label{i4a} \\
\bar\psi(x)&=\bar\psi^{(h)}(x)-\int\limits_{AdS}d^{d+1}y\bar F(y)S(y;x)
\label{i4b}
\end{align}
with:
\begin{align}
\psi^{(h)}(x)&=\int\frac{d^d\vec k}{(2\pi)^d}\cK_{\Delta}(x;\vec k)
\psi^{-(h)}_0(\vec k)= \int\frac{d^d\vec k}{(2\pi)^d}\cK_{-\Delta}(x;\vec k)
\psi^{+(h)}_0(\vec k)
\label{k4a} \\
\bar\psi^{(h)}(x)&=\int\frac{d^d\vec k}{(2\pi)^d}
\bar\psi^{-(h)}_0(\vec k)\bar\cK_{\Delta}(x;\vec k)
= \int\frac{d^d\vec k}{(2\pi)^d}
\bar\psi^{+(h)}_0(\vec k)\bar\cK_{-\Delta}(x;\vec k)
\label{k4b}
\end{align}
where $\cK_{\Delta}$ stands for bulk-boundary propagator defined in the next
Section. In our case the boundary spinors $\psi^{\mp(h)}_0$ couple on the
boundary to operators of dimension $\frac{d+1}2\pm m$ and we call them as
regular and irregular boundary values respectively. We will treat
$\bar\psi^{\pm(h)}_0$ analogously. The case of spinor field possessing regular
boundary conditions are well studied in the context of AdS/CFT correspondence
{see for instance \cite{HS,8}) and we will review their properties in the
next Section.

%%%%%%%%%%%%%%%%%%%%%%%%%%%%%%%%%%%%%%%%%%%%%%%%%%%%%%%%%%%%
\section{Regular Boundary Conditions}

In this Section we will consider the case when the regular boundary conditions
are imposed. The free field solutions $\psi^{(h)}(x)$ and
$\bar\psi^{(h)}(x)$ can be written explicitly as \cite{HS,8,H}:
\begin{align}
\label{10}
\psi^{(h)}(x) & = \int\frac{d^d\vec k}
{(2\pi)^d}\cK_\Delta(kx_0)\psi^{-(h)}_0(\vec k)=
2x_0^{\frac{d+1}{2}}\[K_{m+\frac 12}(kx_0)-i\frac{\hat k}{k}K_{m-\frac
12}(kx_0)\]\(\frac k2\)^{m+\frac 12}\psi^{-(h)}_0(\vec k)\\ \notag
&=\int\frac{d^d\vec k}{(2\pi)^d}\cK_{-\Delta}(kx_0)\psi^{+(h)}_0(\vec k)=
2x_0^{\frac{d+1}{2}}\[K_{\frac 12-m}(kx_0)-i\frac{\hat k}{k}K_{m+\frac
12}(kx_0)\]\(\frac k2\)^{\frac 12-m}\psi^{+(h)}_0(\vec k)
\end{align}
and
\begin{align}
\label{11}
\bar\psi^{(h)}(x) & = \int\frac{d^d\vec k}
{(2\pi)^d}\bar\psi^{-(h)}_0(\vec k)\bar\cK_{\Delta}(kx_0)=
2x_0^{\frac{d+1}{2}}\bar\psi^{-(h)}_0(\vec k)\[K_{m+\frac 12}(kx_0)
+i\frac{\hat k}{k}K_{m-\frac 12}(kx_0)\]\(\frac k2\)^{m+\frac 12}\\ \notag
&=\int\frac{d^d\vec k}{(2\pi)^d}\bar\psi^{+(h)}_0(\vec k)
\bar\cK_{-\Delta}(kx_0)=
2x_0^{\frac{d+1}{2}}\bar\psi^{+(h)}_0(\vec k)\[K_{\frac 12-m}(kx_0)
-i\frac{\hat k}{k}K_{m+\frac 12}(kx_0)\]\(\frac k2\)^{\frac 12-m}
\end{align}
where $K_\a$ are the modified Bessel functions. In the above to obtain
the second equalities we used the on-shell connection between
$\psi^{\pm(h)}_0$ and $\bar\psi^{\pm(h)}_0$ respectively\footnote{The
detailed derivation can be found in \cite{HS,8,H}}:
\begin{align}
&\psi^{+(h)}_0(\vec k)=-i\frac{\hat k}{k}\frac{\G(\frac 12-m)}{\G(\frac
12+m)}\(\frac k2\)^{2m}\psi^{-(h)}_0(\vec k)
\label{12}\\
&\bar\psi^{+(h)}_0(\vec k)=i\bar\psi^{-(h)}_0(\vec k)\frac{\hat k}{k}
\frac{\G(\frac 12-m)}{\G(\frac 12+m)}\(\frac k2\)^{2m}
\label{13}
\end{align}
The bulk-boundary propagator can be easily obtained in the position space by
inverse Fourier transform. The behavior of the free field solutiona in
the limit $x_0\to 0$ can be obtained by using the small argument
expansion of $K_\a$:
\begin{align}
\label{14}
\psi^{(h)}(x)&\overset{x_0\to 0}{\approx} x_0^{\frac d2-1}\psi^{-(h)}_0
(\vec x)+x_0^{\frac d2+1}\psi^{+(h)}_0(\vec x)\\
\label{15}
\bar\psi^{(h)}(x)&\overset{x_0\to 0}{\approx} x_0^{\frac d2-1}
\bar\psi^{-(h)}_0(\vec x)+x_0^{\frac d2+1}\bar\psi^{+(h)}_0(\vec x)
\end{align}

Let us discuss the bulk-bulk Green's function considered in detais for
instance in \cite{OK}. One can relate the spinor propagator to the
scalar Green's function by the ansatz:
$$
S(x;y)=\(\Dir +m\)G_{\frac{d+1}{2}+m}(x;y).
$$
Then the equation (\ref{i3}) for $G$ becomes exactly as for the scalar
Green's function. The explicit solution is found to be:
\begin{align}
\label{g}
&G(x,y) =
 -(x_0y_0)^{\frac{d+1}{2}}\int\frac{d^d\vec k}{(2\pi)^d}
e^{-i\vec k\cdot(\vec x-\vec y)} \\ \notag
&\times\[[K_{\frac 12+m}(kx_0)-i\frac{\hat k}kK_{\frac 12-m}(kx_0)]
\Pi_-[I_{m-\frac 12}(ky_0)-i\frac{\hat k}kI_{\frac 12+m}(ky_0)]
\theta(x_0-y_0)\right.
\\ \notag
&-\left.[I_{m+\frac 12}(kx_0)+i\frac{\hat k}kI_{m-\frac 12}(kx_0)]
\Pi_-[K_{\frac 12-m}(ky_0)+i\frac{\hat k}kK_{\frac 12+m}(ky_0)]
\theta(y_0-x_0)\]\\
&=-\frac{1}{(x_0y_0)^{\frac 12}}\[\bigl(\hat x_0\Pi_--\Pi_+\hat y_0
\bigr)G'_{\frac{d-1}{2}+m}(x;y)+\bigl(\hat x_0\Pi_+-\Pi_-\hat y_0
\bigr)G'_{\frac{d+1}{2}+m}(x;y)\]
\end{align}
where $G_\Delta$ is the standard scalar bulk-bulk propagator, prime
denote derivative with respect to the chordal distance $u=\frac
{(x-y)^2}{2x_0y_0}$ and $\Pi_\pm=\frac 12(I\pm\g_0)$.
From the second line of eq.(\ref{g}) it is straightforward to obtain
the behavior of $S(x;y)$ near the boundary giving:
\begin{align}
\label{19}
S(x;y)\overset{y_0\to 0}{\approx} -y_0^{\frac d2+m}\cK_{\Delta}(\vec x;y);
\qquad
&S(x;y)\overset{x_0\to 0}{\approx} -x_0^{\frac d2+m}\bar\cK_{\Delta}(\vec x;y)
\end{align}
where $\cK_\Delta$ and $\bar\cK_\Delta$ are defined in (\ref{10},\ref{11}).

Let us consider the behavior of the full interacting spinor field in the
limit $x_0\to 0$. In this limit one can write:
\begin{align}
\label{20}
&\psi(x)\overset{x_0\to 0}{\approx} x_0^{\frac d2-m}\psi^-_0(\vec x)
+x_0^{\frac d2+m}\psi^+_0(\vec x)\\ \notag
&\bar\psi(x)\overset{x_0\to 0}{\approx} x_0^{\frac d2-m}\bar\psi^-_0(\vec x)
+x_0^{\frac d2+m}\bar\psi^+_0(\vec x)
\end{align}
Substitution of the asymptotics (\ref{14},\ref{15},\ref{19})
into (\ref{i4a},\ref{i4b}) show
that the interacting part contributes to $\psi^+$ and $\bar\psi^+$ only:
\begin{align}
\label{21}
\psi^-_0(\vec x)&=\psi^{-(h)}_0(\vec x)\\ \notag
\psi^+_0(\vec x)&=\psi^{+(h)}_0(\vec x)+\int_{AdS}d^{d+1}y
\bar\cK_{\Delta}(\vec x;y)F(y)
\\
\label{22}
\bar\psi^-_0(\vec x)&=\bar\psi^{-(h)}_0(\vec x)\\ \notag
\bar\psi^+_0(\vec x)&=\bar\psi^{+(h)}_0(\vec x)+
\int_{AdS}d^{d+1}y\bar F(y)\cK_{\Delta}(\vec x;y)
\end{align}
Let us discuss the boundary term following for instance from the
considerations in \cite{H}. It is based on the requirement to have well
defined variational principle for the action under consideration. Since
the Dirac equations (\ref{eq},\ref{bareq}) are first order differential
equations we cannot fix simultaneously all the components on the boundary
but only half of them, $\psi^+$ or $\psi^-$. The regular boundary
conditions imply that we have fixed $\psi^-,\bar\psi^-$ as a boundary data.
In this case
$\psi^-$($\bar\psi^-$) will serve as a source for the correlation functions
in the boudary theory but $\psi^+$($\bar\psi^+$) will be free to vary. The
variation of the Dirac action in the class of field configurations
(\ref{14},\ref{15}) will produce a surface term contribution:
$$
\d S_D=T_\infty+[0]_{on-shell}
$$
where
\begin{equation}
T_\infty=-\frac 12\int d^d\bx\[\bar\psi^-\(\bx\)\d\psi^+\(\bx\)+
\d\bar\psi^+\(\bx\)\psi^-\(\bx\)\]=-\d S_\infty
\end{equation}
Therefore the full on-shell action with well defined variational principle will
include a boundary term and can be written as:
\begin{align}
\label{a21}
S[\psi^-,\bar\psi^-]&=S_D+S_\infty+S_{int}\\ \notag
&=i\frac{\G(\frac 12-m)}{\G(\frac 12+m)}\int\frac{d^d\vec k}{(2\pi)^d}
\bar\psi^-_0(\vec k)\frac{\hat k}{k}\(\frac k2\)^{2m}\psi^-_0(-\vec k)+
\int\limits_{AdS}d^{d+1}x\,d^{d+1}y\bar F(x)S(x;y)F(y)+S_{int}
\end{align}
Note that the first term in the position space takes the form:
\begin{equation}
S_{cl}^-=-\frac{\G(\frac{d+1}{2}+m)}{\pi^{\frac d2}\G(\frac 12+m)}
\int\limits_{\pa AdS}d^d\vec xd^d\vec y\bar\psi^{-(h)}_0(\vec x)
\frac{\vec\g.(\vec x-\vec y)}{|\vec x-\vec y|^{d+2m+1}}\psi^{-(h)}_0(\vec y)
\label{22a}
\end{equation}
which gives the correct two point correlation function in CFT of
conformal operator with dimension $\D=\frac{d+1}{2}+m$\cite{HS,8,H}. The
other two terms in (\ref{a21}) can be expressed as perturbative series
in terms of $\psi_0$ and $\bar\psi_0$. Since these fields are connected
to the boundary data $\psi^{-(h)}_0$ and $\bar\psi^{-(h)}_0$ (see
eq.(\ref{21},\ref{22})) this would be a perturbative expansion in terms of
$\psi^{-(h)}_0$ and $\bar\psi^{-(h)}_0$. Note that in calculating spinor
exchange one must use for internal lines the Green's function given by
eq.(\ref{g}).

%%%%%%%%%%%%%%%%%%%%%%%%%%%%%%%%%%%%%%%%%%%%%%%%%%%%%%%
\section{Theory from irregular boundary conditions.}

In this Section we will follow the Klebanov and Witten's idea of
taking irregular boundary conditions. In considerations of the
variational principle for spinor field and from the asymptotic behavior
of the field solutions (\ref{14},\ref{15}) it follows that we can choose to fix
$\psi^+$($\bar\psi^+$) instead $\psi^-$($\bar\psi^-$) and the other
components to be free to vary. In the case of scalar field \cite{KW,Wi} it
was shown that the resulting two theories (with regular and irregular
boundary conditions) are connected by a Legendre transform. To show that
it is true also in the case of spinor field let us consider the
expression:
\begin{align}
\label{23}
\frac{\d S[\psi^-,\bar\psi^-]}{\d\psi^-_0(\vec k)}
=-i\frac{\G(\frac 12-m)}{\G(\frac 12+m)}
\bar\psi^-_0(-\vec k)\frac{\hat k}{k}\(\frac k2\)^{2m}+\int\limits_{AdS}
d^{d+1}xd^{d+1}y\bar F(x)S(x;y)\frac{\d F(y)}{\d\psi}\frac{\d\psi}
{\d\psi^-_0(\vec k)} \\ \notag
+\int\limits_{AdS}d^{d+1}x\bar F(x)\frac{\d\psi}{\d\psi^-_0(\vec k)}
\\
\label{24}
\frac{\d S[\psi^-,\bar\psi^-]}{\d\bar\psi^-_0(\vec k)}
=i\frac{\G(\frac 12-m)}{\G(\frac 12+m)}
\frac{\hat k}{k}\(\frac k2\)^{2m}\psi^-_0(-\vec k)+\int\limits_{AdS}
d^{d+1}xd^{d+1}y\frac{\d\bar F(x)}{\d\bar\psi}\frac{\d\bar\psi}
{\d\bar\psi^-_0(\vec k)}S(x;y)F(y)\\ \notag
+\int\limits_{AdS}d^{d+1}x\frac{\d\bar\psi}{\d\bar\psi^-_0(\vec k)}F(y)
\end{align}
Using (\ref{12},\ref{13}) one can see that the first tems in
(\ref{23},\ref{24}) are
$\bar\psi^+_0$ and $\psi^+_0$ respectively. We will subsitute $\frac
{\d\psi}{\d\psi^-_0}$ and $\frac{\d\bar\psi}{\d\bar\psi^-_0}$ in the
last terms in (\ref{23},\ref{24}) by expressions found from
(\ref{i4a},\ref{i4b}):
\begin{align}
\label{25}
\frac{\d\psi}{\d\psi^-_0(\vec k)}=\cK_{\Delta}(x,-\vec
k)-\int\limits_{AdS}d^{d+1}yS(x;y)\frac{\d F(y)}{\d\psi}\frac
{\d\psi}{\d\psi^-_0(\vec k)}
\\
\frac{\d\bar\psi}{\d\bar\psi^-_0(\vec k)}=\bar\cK_{\Delta}(-\vec k,x)
-\int\limits_{AdS}d^{d+1}y
\frac{\d\bar F(y)}{\d\psi}\frac{\d\bar\psi}{\d\bar\psi^-_0(\vec k)}S(x;y)
\label{26}
\end{align}
The result is:
\begin{align}
\frac{\d S[\psi^-,\bar\psi^-]}{\d\psi^-_0(\vec k)}
=\bar\psi^{+(h)}_0(-\vec k)+
\int\limits_{AdS}d^{d+1}x\bar F(x)\cK_{\Delta}(-\vec
k,x)=\bar\psi^+_0(-\vec k)
\label{27}\\
\frac{\d S[\psi^-,\bar\psi^-]}{\d\bar\psi^-_0(\vec k)}
=\psi^{+(h)}_0(-\vec k)+
\int\limits_{AdS}d^{d+1}x\bar\cK_{\Delta}(x,-\vec k) F(x)=\psi^+_0(-\vec k)
\label{28}
\end{align}
and this relations are valid to any order in perturbation theory.
Therefore, $\psi^-_0$ and $\bar\psi^+_0$ can be considered as conjugated and
the same is true for the pair ($\bar\psi^-_0,\psi^+_0$).

Obviously the functional defined by a Legendre transform:
\begin{equation}
J[\psi^-,\psi^+;\bar\psi^-,\bar\psi^+]=S[\psi^-,\bar\psi^-]-
\int d^dx\[\bar\psi^+\psi^-+\bar\psi^-\psi^+\]
\label{29}
\end{equation}
possess a minimum with respect to variation of $\psi^-$ and
$\bar\psi^-$. At the extremal point, the functional (\ref{29}) is
expressed in terms of $\psi^+$ and $\bar\psi^+$ only. As it is clear
from the preceeding discussion, $\psi^+$ and $\bar\psi^+$ are related to
the choice of irregular boundary conditions and Klebanov and Witten
\cite{KW} proposed to consider AdS/CFT correspondence principle based on
the use of $J[\psi^+,\bar\psi^+]$ instead of $S[\psi^-,\bar\psi^-]$.

Let us compute the minimum of $J$ and to see what kind of theory is
induced on the boundary. Subsitution of $S[\psi^-,\bar\psi^-]$ from
(\ref{a21}) in (\ref{29}) gives:
\begin{align}
\label{30}
J\[\psi^+,\bar\psi^+\]=&-\frac 12\int d^d\vec k\[\bar\psi^-(-\vec k)
\psi^+(\vec k)+\bar\psi^+(\vec k)\psi^-(-\vec k)\] \\ \notag
&+\int\limits_{AdS}d^{d+1}xd^{d+1}y\bar F(x)S(x;y)F(y)+S_{int}
\end{align}
Using the eqs. (\ref{12}) and (\ref{13}) and the relation:
\begin{align}
\label{a30}
\psi^{+(h)}_0(-\vec k)&=\psi^+_0(-\vec k)
-\int_{AdS}d^{d+1}y\bar\cK_{\Delta}(y,\vec k) F(y) \\
\label{b30}
\bar\psi^{+(h)}_0(-\vec k)&=\bar\psi^+_0(-\vec k)
-\int_{AdS}d^{d+1}y\bar F(x)\cK_{\Delta}(x,\vec k)
\end{align}
one can subsitute $\psi^-$ and $\bar\psi^-$ into the first term in eq.
(\ref{30}). The resulting functional takes the form:
\begin{align}
\label{31}
J\[\psi^+,\bar\psi^+\]=
&i\frac{\G(\frac 12+m)}{\G(\frac 12-m)}\int\frac{d^d\vec k}{(2\pi)^d}
\bar\psi^+(\vec k)\frac{\hat k}{k}\(\frac k2\)^{-2m}\psi^+(-\vec k)
+\int_{AdS}d^{d+1}xd^{d+1}y\bar F(x)S(x;y)F(y) \\ \notag
&+\int_{AdS}d^{d+1}xd^{d+1}y
\bar F(x)\int\frac{d^d\vec k}{(2\pi)^d}e^{-\vec k\cdot(\vec x-\vec y)}
\cK(x,\vec k)\Pi_-\bar\cK_{-\Delta}(y,\vec k)F(y)+S_{int}
 \\
\label{32}
=&S_{cl}^+[\psi^+,\bar\psi^+]+S_{int}\\ \notag
&+\int\limits_{AdS}d^{d+1}x\bar F(x)
\[S(x;y)+
\int d^{d+1}y\frac{d^d\vec k}{(2\pi)^d}e^{-i\vec k\cdot(\vec x-\vec y)}
\cK_\Delta(x,\vec k)\Pi_-
\bar\cK_{-\Delta}(y,\vec k)\]F(y)
\end{align}
where $\bar\cK_{-\Delta}$ means that in $\bar\cK_{\Delta}$ $m$ is replaced
by $-m$ (see eq.(\ref{11})).

The first term in (\ref{32}) can be easily Fourier transformed to the
position space:
\begin{equation}
S_{cl}^+=-\frac{\G(\frac{d+1}{2}-m)}{\pi^{\frac d2}\G(\frac 12-m)}
\int d^dx\,d^dy\,\bar\psi^+(\vec x)
\frac{\vec\g.(\vec x-\vec y)}{|\vec x-\vec y|^{d+1-2m}}
\label{33}
\end{equation}
which coincide with the two point correlation function of an operator in
the boundary CFT with dimension $\D_-=\frac{d+1}{2}-m$. Therefore,
Klebanov and Witten's idea to define AdS/CFT correspondence by taking
irregular boundary conditions will be formulated in our case by the
formula:
\begin{equation}
e^{-J[\psi^+,\bar\psi^+]}=\langle
exp\int\[\bar{\mathcal O}\psi^++\bar\psi^+{\mathcal O}\]\rangle
\label{34}
\end{equation}
Note that in contrast to the scalar case, due to the specific
asymptotics in the spinor case there is no additional multiplier infront
of the exponent on the right hand side of (\ref{34}). Using such a
formulation one can see that the Green's function that must be used for
the internal lines in calculating the higher order graphs get modified and is
defined by the expression in the square brackets multplying $\bar F(x)$
and $F(y)$ on the right hand side of (\ref{32}):
\begin{equation}
\tilde S(x;y)=S(x;y)+\int\frac{d^d\vec k}{(2\pi)^d}
e^{-i\vec k\cdot(\vec x-\vec y)}
\cK_{\Delta}(x,\vec k)\Pi_-\bar\cK_{-\Delta}(y,\vec k)
\label{35}
\end{equation}

Let us analyse the modified Green's function $\tilde S(x;y)$. For this
purpose we will use for $I_\a$ part of $S(x;y)$ (\ref{g}) the following
relations:
\begin{align}
&I_{m+\frac 12}(z)=I_{-m-\frac 12}(z)-\frac{2K_{\frac 12+m}(z)}
{\G(\frac 12-m)\G(\frac 12+m)};\qquad
&I_{m-\frac 12}(z)=I_{\frac 12-m}(z)+\frac{2K_{\frac 12-m}(z)}
{\G(\frac 12-m)\G(\frac 12+m)}
\label{36}
\end{align}
and the property of the Macdonald function $K_\a(z)=-K_{-\a}(z)$.
Insearting (\ref{36}) into (\ref{g}) we find the expression for the
Green's fuction:
\begin{align}
\label{37}
&G(x,y) =- \int\frac{d^d\vec k}{(2\pi)^d}e^{-i\vec k\cdot(\vec x-\vec y)}
\cK_{\Delta}(x,\vec k)\Pi_-\bar\cK_{-\Delta}(\vec k,y)
\\ \notag
& -(x_0y_0)^{\frac{d+1}{2}}\int\frac{d^d\vec k}{(2\pi)^d}
e^{-i\vec k\cdot(\vec x-\vec y)}
\[[K_{\frac 12-m}(kx_0)-i\frac{\hat k}kK_{\frac 12+m}(kx_0)]
\Pi_-[I_{-m-\frac 12}(ky_0)-i\frac{\hat k}kI_{\frac 12-m}(ky_0)]\right.
\\ \notag
&\times\theta(x_0-y_0)
-\left.[I_{-m+\frac 12}(kx_0)+i\frac{\hat k}kI_{-m-\frac 12}(kx_0)]
\Pi_-[K_{\frac 12+m}(ky_0)+i\frac{\hat k}kK_{\frac 12-m}(ky_0)]
\theta(y_0-x_0)\]\\ \notag
&\qquad =- \int\frac{d^d\vec k}{(2\pi)^d}e^{-i\vec k\cdot(\vec x-\vec y)}
\cK_{\Delta}(x,\vec k)\Pi_-\bar\cK_{-\Delta}(\vec k,y)+\tilde S(x;y).
\end{align}
where:
\begin{align}
\tilde S(x;y)=&-(x_0y_0)^{\frac{d+1}{2}}\int\frac{d^d\vec k}{(2\pi)^d}
e^{-i\vec k\cdot(\vec x-\vec y)}\\ \notag
&\times\[[K_{\frac 12-m}(kx_0)-i\frac{\hat k}kK_{\frac 12+m}(kx_0)]
\Pi_-[I_{-m-\frac 12}(ky_0)-i\frac{\hat k}kI_{\frac 12-m}(ky_0)]\theta(x_0-y_0)
\right.
\\ \notag
&-\left.[I_{-m+\frac 12}(kx_0)+i\frac{\hat k}kI_{-m-\frac 12}(kx_0)]
\Pi_-[K_{\frac 12+m}(ky_0)+i\frac{\hat k}kK_{\frac 12-m}(ky_0)]
\theta(y_0-x_0)\]
\end{align}
Therefore, as in the bosonic case, the modified Green's function $\tilde
S(x;y)$ is obtained by replacing $m$ with $-m$. The new action that must
be used in the AdS/CFT correspondence is:
\begin{equation}
J\[\psi^+,\bar\psi^+\]=S_{cl}^+\[\psi^+,\bar\psi^+\]+
\int_{AdS}d^{d+1}xd^{d+1}y\bar F(x)\tilde S(x;y)F(y)+S_{int}
\label{38}
\end{equation}
where $S_{cl}^+$ defines the two point correlation function (\ref{33}).
The representation of $\psi (x)$ and $\bar\psi (x)$ in terms of $\psi^+$
and $\bar\psi^+$:
\begin{align}
\psi(x)&=\int d^dy\cK_{-\Delta}(x,\vec y)\psi^+_0(\vec y)-
\int_{AdS}d^{d+1}y\tilde S(x;y)F(y)
\label{39}\\
\bar\psi(x)&=\int d^dy\bar\cK_{-\Delta}(x,\vec y)\psi^+_0(\vec y)-
\int_{AdS}d^{d+1}y\bar F(y)\tilde S(x;y)
\label{40}
\end{align}
and the asymptotics:
\begin{align}
\tilde S(x;y)\overset{y_0\to 0}{\approx} -y_0^{\frac d2-m}
\cK_{-\Delta}(\vec x;y),
&\qquad \tilde S(x;y)\overset{x_0\to 0}{\approx} -x_0^{\frac d2-m}
\bar\cK_{-\Delta}(\vec x;y)
\label{41}
\end{align}
show that the interaction contributes to $\psi^-$ and $\bar\psi^-$
only. In the calculation of the spinor exchange for internal
lines one must use $\tilde S$. Repeating all the considerations, but now
with $J[\psi^+,\bar\psi^+]$ instead of $S[\psi^-,\bar\psi^-]$ one can
find that:
\begin{equation}
\frac{\d J}{\d\psi^+}=\bar\psi^-;\qquad \frac{\d J}{\d\bar\psi^+}=\psi^-
\end{equation}

Let us summarize our considerations in the next Section.
%%%%%%%%%%%%%%%%%%%%%%%%%%%%%%%%%%%%%%%%%%%%%%%%%%%%

\ \\

{\bf{\large Conclusions}}
\ \\
In this acticle we generalized the Klebanov and Witten idea \cite{KW}
(generalized to the interacting case in \cite{Wi}) to the spinor case.
In \cite{KW} it was prposed to consider AdS/CFT correspondence for
bosonic case based on a Legendre transformed action. As in the bosonic
case, there are two possible choices of boundary conditions for the
spinor part of the action of AdS supergarvity - regular and irregular.
The first of them couples to the spinors ($\psi^-,\bar\psi^-$) operators
with dimension ($\frac{d+1}{2}+m$) while the second one couples to the
spinors ($\psi^+,\bar\psi^+$) operators with dimension
($\frac{d+1}{2}-m$). The generating functional of correlation functions
are related by a Legendre transform and this is true to all orders in
perturbation theory.

The Green's function (\ref{g}) that must be used for internal lines in
calculating higher order graphs is replaced by a modified Green's
function $\tilde S(x;y)$ which explicit form is given by eq.(\ref{41}).

It would be interesting to compare the results for spinor exchange
and the "transfer functions" in both theories \cite{OK,Bala}.
Work in these topics is in progress.
We hope to present analogous considerations for the case of vector and
Rarita-Schwinger fields in the near future \cite{Ra}.

\ \\

{\bf Acknowledgments}

I am grateful to N.I.Karchev and M.Stanishkov for comments and critical
reading the manuscript.

\ \\


\begin{thebibliography}{99}
\bibitem{Malda} J. Maldacena, "The Large N Limit of Superconformal Field
Theories and Supergravity", Adv. Theor. Math. Phys. 2
(1998) 231, hep-th/9711200.
\bibitem{gkp} S. S. Gubser,I. R. Klebanov
and A. M. Polyakov, Gauge Theory Correlators from Non-critical String
Theory, Phys. Lett. B428 (1998) 105, hep-th/9802109.
\bibitem{w} E. Witten, Anti De Sitter Space
and Holography, Adv. Theor. Math. Phys. 2 (1998) 253, hep-th/9802150.
\bibitem{KW} I.Klebanov and E.Witten, AdS/CFT Correspondence and
Symmetry Breaking, hep-th/9905104.
\bibitem{Wi} W.M\"uck and K.S.Viswanathan, Regular and Irregular
Boundary Conditions in the AdS/CFT Correspondence, hep-th/9906155.
\bibitem{5} W. M\"{u}ck and K. S. Viswanathan,
Conformal Field Theory Correlators from Classical Scalar Field Theory
on $AdS_{d+1}$, Phys. Rev. D58 (1998) 041901, hep-th/9804035.
\bibitem{6} D. Z. Freedman, S. D. Mathur, A. Matusis and
L. Rastelli, Correlation functions in the $CFT_d/AdS_{d+1}$
correspondence, hep-th/9804058.
\bibitem{8} W. M\"{u}ck and K. S. Viswanathan,
Conformal Field Theory Correlators from Classical Field Theory on Anti-de
Sitter Space II. Vector and Spinor Fields,
Phys. Rev. D58 (1998) 106006, hep-th/9805145.
\bibitem{9} G. Chalmers, H. Nastase, K. Schalm and
R. Siebelink, R-Current Correlators in ${\cal N}=4$ Super Yang-Mills
Theory from Anti-de Sitter Supergravity, hep-th/9805105.
\bibitem{HS} M. Henningson and K. Sfetsos,
Spinors and the AdS/CFT correspondence, Phys. Lett. B431 (1998) 63,
hep-th/9803251.
\bibitem{12} G. E. Arutyunov and S. A. Frolov, On the origin of
supergravity boundary terms in the AdS/CFT correspondence,
hep-th/9806216.
\bibitem{14} G. E. Arutyunov and S. A. Frolov, Antisymmetric tensor
field on $AdS_5$, hep-th/9807046.
\bibitem{15} W. S. l'Yi, Generating functionals of correlation
functions of massive vector field in AdS/CFT correspondence,
hep-th/9809132.
\bibitem{C}  S. Corley, The massless gravitino and the AdS/CFT
correspondence, hep-th /9808184.
\bibitem{A} A. Volovich, Rarita-Schwinger Field in the AdS/CFT
 Correspondence, hep-th/9809009.
\bibitem{kr} A. S. Koshelev and O. A. Rytchkov , Note on the Rarita-Schwinger
Field in the AdS/CFT Correspondence, hep-th/9812238.
\bibitem{Rash} R.C.Rashkov, Note on the boundary terms in AdS/CFT
correspondence for Rarita-Schwinger field, hep-th/9904098.
\bibitem{19} A. Ghezelbash, K. Kaviani, S. Parvizi and A. Fatollahi,
Interacting Spinors - Scalars and ADS/CFT Correspondence,
Phys. Lett. B435 (1998) 291, hep-th/9805162.
\bibitem{H} M. Henneaux, Boundary terms in the AdS/CFT correspondence for
spinor fields, hep-th/9902137.
\bibitem{OK} T.Kawano and K.Okuyama, Spinor Exchange in $AdS_{d+1}/CFT$,
hep-th/9905130.
\bibitem{P} J. Polchinski, S-matrices from AdS spacetime, hep-th/9901076.
\bibitem{Sus1} L. Suskind, Holography in the flat-space limit,
hep-th/9901079.
\bibitem{Bala}  V. Balasubramanian, S. Giddings, A. Lawrence,
What Do CFTs Tell Us About Anti-de Sitter Spacetimes?, hep-th/9902052.
\bibitem{wis} P.Matlock and K.S.Wisanathan, The AdS/CFT Correpondence
for the Massive Rarita-Schwinger field, hep-th/9906077
\bibitem{Vlado}  V.K.Dobrev, Intertwining Operator Realization of the  AdS/CFT
Correspondence, hep-th/9812194.
\bibitem{OM} O.Aharovy, S.S.Gubser, J.Maldacena, H.Ooguri, Y.Oz,
Lagre $N$ Theories, String theory and Gravity, hep-th/9905111.
\bibitem{Ra} R.C.Rashkov, In preparation.
\end{thebibliography}
\end{document}